\documentclass[aps,prd,twocolumn,showpacs,preprintnumbers,amsmath,amssymb,nofootinbib]{revtex4-1}
\usepackage{amsmath,amsfonts,euscript,amssymb}
\usepackage{epsfig}
\DeclareSymbolFont{bfitletters}{OML}{cmm}{bx}{it}
\DeclareSymbolFont{bfitoperators}   {OT1}{cmr} {m}{n}
\DeclareMathSymbol{\bfitomega}{\mathord}{bfitletters}{"21}

\newcommand{\be}{\begin{equation}}
\newcommand{\ee}{\end{equation}}
\newcommand{\bea}{\begin{eqnarray}}
\newcommand{\eea}{\end{eqnarray}}
\begin{document}

\title{Exact solutions of Friedmann equation for supernovae data}

\author{A. E. Pavlov$^{1,2}$}
\affiliation{${}^1$Bogoliubov~Laboratory~for~Theoretical~Physics,~Joint~Institute~of~Nuclear~Research,
~Joliot-Curie~str.~6,~Dubna,~141980,~Russia \\
$^{2}$Institute of Mechanics and Energetics, Russian State Agrarian University, Timiryazevskaya, 49,
Moscow 127550, Russia\\
alexpavlov60@mail.ru}

\begin{abstract}
An intrinsic time of homogeneous models is global. The Friedmann equation by its sense ties time intervals.
Exact solutions of the Friedmann equation in Standard cosmology and Conformal cosmology are presented.
Theoretical curves interpolated the Hubble diagram on latest supernovae are expressed in analytical form. 
The class of functions in which the concordance model is described is Weierstrass meromorphic functions. 
The Standard cosmological model and Conformal one fit the modern Hubble diagram equivalently. 
However, the physical interpretation of the modern data from concepts of the Conformal cosmology is simpler, so is preferable.
\end{abstract}



\maketitle

\section{Introduction}

The supernovae type Ia are used as standard candles to test cosmological models. 
Recent observations of the supernovae have led cosmologists to conclusion of the Universe filled with dust and 
mysterious dark energy in frame of Standard cosmology \cite{Riess2004}.
Recent cosmological data on expanding Universe challenge cosmologists in insight of Einstein's gravitation.
To explain a reason of the Universe's acceleration the significant efforts have been applied (see, for example, \cite{Bor, Sz}).

The Conformal cosmological model \cite{PP} allows us to describe the supernova data without Lambda term. 
The evolution of the lengths in the Standard cosmology is replaced by the evolution of the masses in the Conformal cosmology. 
It allows to hope for solving chronic problems accumulated in the Standard cosmology. 
Solutions of the Friedmann differential equation belong to a class of Weierstrass meromorphic functions. 
Thus, it is natural to use them for comparison predictions of these two approaches. 
The paper presents a continuation of the article on intrinsic time in Geometrodynamics \cite{Geom}.

\section{Friedmann equation in Classical cosmology}

A global time exists in homogeneous cosmological models (see, for example, papers \cite{Kasner,Misner}).
The conformal metric $(\tilde\gamma_{ij})$ \cite{PP} for three-dimensional sphere in spherical coordinates
$(\chi, \theta, \varphi)$ is defined via the first quadratic form
\begin{equation}\label{back}
a_0^2\left[d\chi^2+\sin^2\chi (d\theta^2+\sin^2\theta d\varphi^2)\right].
\end{equation}
Here $a_0$ is a modern value of the Universe's scale.
For a pseudosphere in (\ref{back}) instead of $\sin\chi$ one should take $\sinh\chi$, and for a flat space
one should take $\chi.$ The intrinsic time $D$ is defined with minus as logarithm of ratio of scales
$$D=-\ln\left(\frac{a(t)}{a_0}\right).$$

In the Standard cosmological model the Friedmann equation is used for fitting SNe Ia data. It
ties the intrinsic intervals time with the coordinate time one
\begin{equation}\label{Friedmannclass}
\left(\frac{dD}{dt}\right)^2\equiv
\left(\frac{\dot{a}}{a}\right)^2=H_0^2\left[\Omega_{\rm M}\left(\frac{a_0}{a}\right)^3+\Omega_{\Lambda}\right].
\end{equation}
Three cosmological parameters favor for modern astronomical observations
$$H_0=h\cdot 10^5 m/s/Mpc,\quad h=0.72\pm 0.08$$
-- Hubble constant,
$$\Omega_\Lambda = 0.72,\quad \Omega_{\rm M}=0.28$$
-- partial densities. Here $\Omega_{\rm M}$ is the baryonic density parameter, $\Omega_\Lambda$ 
is the density parameter corresponding to $\Lambda$-term, constrained with $\Omega_{\rm M}+\Omega_\Lambda =1.$

The solution of the Friedmann equation (\ref{Friedmannclass}) is presented in analytical form
\begin{equation}\label{classicalsolution}
a(t)=a_0\sqrt[3]{\frac{\Omega_{\rm M}}{\Omega_\Lambda}}
\left[{\rm sinh}\left(\frac{3}{2}\sqrt{\Omega_\Lambda}H_0 t)\right)\right]^{2/3}.
\end{equation}
Here $a(t)$ is a scale of the model, $a_0=1$ is its modern value.
The second derivative of the scale factor is
\be
\ddot{a}=\frac{H_0^2 a_0}{2}\left[2\Omega_\Lambda\left(\frac{a}{a_0}\right)-\Omega_{\rm M}\left(\frac{a_0}{a}\right)^2\right].
\ee
In the modern epoch the Universe expands with acceleration, because $2\Omega_\Lambda >\Omega_{\rm M};$
in the past, its acceleration is negative $\ddot{a}<0.$ 
This change of sign of the acceleration without clear physical reason puzzles researchers. 
From the solution (\ref{classicalsolution}), if one puts
$$\frac{a(t)}{a_0}=\frac{1}{1+z},$$
the {\it age -- redshift relation} is followed
\be
H_0t=\frac{2}{3\sqrt{\Omega_\Lambda}}{\rm Arcsinh}
\left(\sqrt{\frac{\Omega_\Lambda}{\Omega_M}}\frac{1}{(1+z)^{3/2}}\right).\label{ageredshiftclassic}
\ee
The age $t_0$ of the modern Universe is able to be obtained by taking $z=0$ in (\ref{ageredshiftclassic})
\begin{equation}
t_0=\frac{2}{3\sqrt{\Omega_\Lambda}}\frac{1}{H_0}{\rm Arcsinh}\sqrt{\frac{\Omega_\Lambda}{\Omega_{\rm M}}}.
\end{equation}

Since for light
$$ds^2=-c^2dt^2+a^2(t)dr^2=0,\quad cdt=-a(t)dr,$$
we have, denoting $x\equiv a/a_0$,
\be\label{int}
-a_0 r=c\int\frac{dt}{x}=c\int\frac{dx}{x}\frac{1}{dx/dt}.
\ee
Rewriting the Friedmann equation (\ref{Friedmannclass}), one obtains a quadrature
\be\label{Friedmannx}
\frac{dx}{dt}=H_0\sqrt{\Omega_{\rm M}/x+\Omega_\Lambda x^2}.
\ee
Substituting the derivative (\ref{Friedmannx}) into (\ref{int}), we get the integral
\be\label{auxiliary}
H_0 r=\frac{c}{\sqrt{\Omega_\Lambda}}\int\limits_{1/(1+z)}^1\frac{dx}{\sqrt{x^4+ 4a_3 x}},
\ee
where we denoted a ratio as
$$4a_3\equiv \frac{\Omega_{\rm M}}{\Omega_\Lambda}.$$
Then we introduce a new variable $y$ by the following substitution
\be\label{xy}
\sqrt{x^4+4a_3 x}\equiv x^2-2y.
\ee
Raising both sides of this equation in square, we get
\be\label{xysquare}
a_3 x=-x^2 y+y^2.
\ee
Differentials of both sides of the equality (\ref{xysquare}) can de expressed in the form:
$$\frac{dx}{x^2-2y}=-\frac{dy}{2xy+a_3}.$$
Utilizing the equality (\ref{xy}), one can rewrite it
\be\label{du}
\frac{dx}{\sqrt{x^4+4a_3x}}=-\frac{dy}{2xy+a_3}.
\ee
Then, we take the expression from the equation (\ref{xysquare})
$$2xy+a_3=\pm\sqrt{4y^3+a_3^2},$$
where a sign plus if
$$0\le x\le\sqrt[3]{\frac{a_3}{2}}=\frac{1}{2}\sqrt[3]{\frac{\Omega_{\rm M}}{\Omega_\Lambda}},$$
and a sign minus if
$$\sqrt[3]{\frac{a_3}{2}}=\frac{1}{2}\sqrt[3]{\frac{\Omega_{\rm M}}{\Omega_\Lambda}}\le x\le 1,$$
and substitute it into the right hand of the differential equation (\ref{du}).
The equation takes the
following form
\bea
&&\frac{dx}{\sqrt{x^4+4a_3 x}}=\mp\frac{dy}{\sqrt{4y^3+a_3^2}}\equiv\nonumber\\
&\equiv&\mp\frac{dy}{2\sqrt{(y-e_1)(y-e_2)(y-e_3)}},\label{W}
\eea
where
$$y\equiv\frac{1}{2}\left(x^2-\sqrt{x^4+4a_3 x}\right),$$
with three roots:
\bea
e_1&\equiv& \frac{1}{8}\left(\frac{\Omega_{\rm M}}{\Omega_\Lambda}\right)^{2/3}(1+\imath\sqrt{3}),\quad
e_2\equiv -\frac{1}{4}\left(\frac{\Omega_{\rm M}}{\Omega_\Lambda}\right)^{2/3},\nonumber\\
e_3&\equiv& \frac{1}{8}\left(\frac{\Omega_{\rm M}}{\Omega_\Lambda}\right)^{2/3}(1-\imath\sqrt{3}).\label{e1e2e3}
\eea
The integral (\ref{auxiliary}) for the interval $\sqrt[3]{a_3/2}\le x\le 1$, corresponding to
\begin{equation}\label{interval}
0\le z\le 2\sqrt[3]{\frac{\Omega_\Lambda}{\Omega_{\rm M}}}\approx 1.74,
\end{equation}
gives the {\it coordinate distance  - redshift relation} in integral form
\bea
&&H_0 r=\nonumber\\
&=&\frac{c}{\sqrt{\Omega_\Lambda}}\int\limits_{[1-\sqrt{1+4a_3(1+z)^3}]/(2(1+z)^2)}^\infty
\frac{dy}{\sqrt{4y^3+a_3^2}}-\nonumber\\
&-&\frac{c}{\sqrt{\Omega_\Lambda}}\int\limits_{(1-\sqrt{1+4a_3})/2}^\infty
\frac{dy}{\sqrt{4y^3+a_3^2}}.\label{WP}
\eea
The interval considered in (\ref{interval}) covers the modern cosmological observations one \cite{Riess2004} 
up to the right latest achieved redshift limit $z\sim 1.7$.

The integrals in (\ref{WP}) are expressed with use of inverse Weierstrass $\wp$-function \cite{Whittaker}
\bea
H_0 r&=&-\frac{c}{\sqrt{\Omega_\Lambda}}\wp^{-1}
\left[\frac{1-\sqrt{1+\Omega_{\rm M}/\Omega_\Lambda(1+z)^3}}{2(1+z)^2}\right]+\nonumber\\
&+&\frac{c}{\sqrt{\Omega_\Lambda}}
\wp^{-1}\left[\frac{(1-\sqrt{1+\Omega_{\rm M}/\Omega_\Lambda})}{2}\right].\label{relationP}
\eea
The invariants of the Weierstrass functions are
$$g_2=0,\qquad g_3=-a_3^2=-\left(\frac{\Omega_{\rm M}}{4\Omega_\Lambda}\right)^2;$$
the discriminant is negative
$$\Delta\equiv g_2^3-27g_3^2<0.$$

Let us rewrite the relation (\ref{relationP}) in implicit form between the variables with use of $\wp$-function
$$\wp u=\frac{1-\sqrt{1+\Omega_{\rm M}/\Omega_\Lambda (1+z)^3}}{2(1+z)^2},$$
where
$$u\equiv \frac{1}{c}\sqrt{\Omega_\Lambda}H_0 r-\wp^{-1}\left(\frac{1-\sqrt{1+\Omega_{\rm M}/\Omega_\Lambda}}{2}\right),$$
The Weierstrass $\wp$-function can be expressed through an elliptic Jacobi cosine function \cite{Whittaker}
\be\label{cnfrac}
\wp u=e_2+H\frac{1+{\rm cn} \left(2\sqrt{H}u\right)}{1-{\rm cn}\left(2\sqrt{H}u\right)},
\ee
where, the roots from (\ref{e1e2e3}) are presented in the form
$$e_1=m+\imath n,\quad e_2=-2m,\quad e_3=m-\imath n,$$
therefore
$$m\equiv \frac{1}{8}\left(\frac{\Omega_{\rm M}}{\Omega_\Lambda}\right)^{2/3},\quad
n\equiv\frac{\sqrt{3}}{8}\left(\frac{\Omega_{\rm M}}{\Omega_\Lambda}\right)^{2/3},$$
and $H$ is calculated according to the rule
$$H\equiv\sqrt{9m^2+n^2}=\frac{\sqrt{3}}{4}\left(\frac{\Omega_{\rm M}}{\Omega_\Lambda}\right)^{2/3}.$$
Then, from (\ref{cnfrac}) we obtain an implicit dependence between the variables, using Jacobi cosine function
\be\label{cnJacobi}
{\rm cn} \left[\sqrt[4]{3}(\Omega_{\rm M}/\Omega_\Lambda)^{1/3}u\right]=\frac{f(z)-1}{f(z)+1},
\ee
where we introduced the function of redshift
\bea
&&f(z)\equiv \nonumber\\
&\equiv&\frac{2}{\sqrt{3}}\left(\frac{\Omega_\Lambda}{\Omega_{\rm M}}\right)^{2/3}
\frac{1-\sqrt{1+\Omega_{\rm M}/\Omega_\Lambda(1+z)^3}}{(1+z)^2}+\frac{1}{\sqrt{3}}.\nonumber
\eea
The modulo of the elliptic function (\ref{cnJacobi}) is obtained by the following rule \cite{Whittaker}:
$$k\equiv\sqrt{\frac{1}{2}-\frac{3e_2}{H}}=\sqrt{\frac{1}{2}+\sqrt{3}}.$$

Claudio Ptolemy classified the stars visible to the naked eye into six classes according to their brightness.
The magnitude scale is a logarithmic scale, so that a difference of 5
magnitudes corresponds to a factor of 100 in luminosity. The absolute magnitude $M$ and the apparent magnitude
$m$ of an object are defined as
\begin{eqnarray}
M&\equiv& -\frac{5}{2} {\rm lg} \frac{L}{L_0},\nonumber\\
m&\equiv& -\frac{5}{2} {\rm lg} \frac{l}{l_0},\nonumber
\end{eqnarray}
where $L_0$ and $l_0$ are reference luminosities. In astronomy, the radiated power $L$ of a star or a galaxy,
is called its absolute luminosity. The flux density $l$ is called its apparent luminosity. In Euclidean
geometry these are related as
$$l=\frac{L}{4\pi d^2},$$
where $d$ is our distance to the object. Thus one defines the {\it luminosity distance} $d_L$ of an object as
\begin{equation}\label{dL}
d_L\equiv\sqrt{\frac{L}{4\pi l}}.
\end{equation}
In Friedmann -- Robertson -- Walker cosmology the absolute luminosity
\begin{equation}\label{L}
L=\frac{N_\gamma E_{\rm em}}{t_{\rm em}},
\end{equation}
where $N_\gamma$ is a number of photons emitted, $E_{\rm em}$ is their average energy,
$t_{\rm em}$ is emission time. The apparent luminosity is expressed as
\begin{equation}\label{l}
l=\frac{N_\gamma E_{\rm abs}}{t_{\rm abs}A},
\end{equation}
where $E_{\rm abs}$ is their average energy, and
$$A=4\pi a_0^2 r^2$$
is an area of the sphere around a star. The number of photons is conserved, but their energy is redshifted,
\begin{equation}\label{Eabs}
E_{\rm abs}=\frac{E_{\rm em}}{1+z}.
\end{equation}
The times are connected by the relation
\begin{equation}\label{tabs}
t_{\rm abs}=(1+z)t_{\rm em}.
\end{equation}
Then, with use of (\ref{Eabs}), (\ref{tabs}), the apparent luminosity (\ref{l}) 
can be presented via the absolute luminosity (\ref{L}) as
$$l=\frac{N_\gamma E_{\rm em}}{t_{\rm em}}
\frac{1}{(1+z)^2}\frac{1}{4\pi a_0^2 r^2}=\frac{1}{(1+z)^2}\frac{L}{4\pi a_0^2 r^2}.$$
From here, the formula for luminosity distance (\ref{dL}) is obtained
\begin{equation}\label{dLs}
d_L (z)_{SC}=(1+z)a_0 r.
\end{equation}

Substituting the formula for coordinate distance (\ref{relationP}) into (\ref{dLs}), 
we obtain the analytical expression for the luminosity distance
\bea
&&d_L(z)_{SC}=\nonumber\\
&&=\frac{c(1+z)}{H_0\sqrt{\Omega_\Lambda}}\left(
\wp^{-1}\left[\frac{(1-\sqrt{1+\Omega_{\rm M}/\Omega_\Lambda})}{2}\right]-\right.\nonumber\\
&-&\left.\wp^{-1}\left[\frac{1-\sqrt{1+\Omega_{\rm M}/\Omega_\Lambda(1+z)^3}}{2(1+z)^2}\right]\right).\nonumber
\eea

The modern observational cosmology is based on the Hubble diagram.
{\it The effective magnitude -- redshift relation}
\begin{equation}\label{mMcl}
m(z)-M=5{\rm lg} [d_L(z)_{SC}]+{\cal M},
\end{equation}
is used to test cosmological theories ($d_L$ in units of megaparsecs) \cite{Riess2004}.
Here $m(z)$ is an observed magnitude, $M$ is the absolute magnitude,
and ${\cal M}=25$ is a constant.

\section{Friedmann equation in Conformal cosmology}

The fit of Conformal cosmological model with $\Omega_{\rm rigid} = 0.755,$ $\Omega_{\rm M}=0.245$
is the same quality approximation
as the fit of the Standard cosmological model with $\Omega_\Lambda = 0.72,$ $\Omega_{\rm M}=0.28$,
constrained with $\Omega_{\rm rigid}+\Omega_\Lambda =1$ \cite{PZakh}.
The parameter $\Omega_{\rm rigid}$ corresponds to a rigid state, where the energy density coincides with the pressure $p=\rho$ \cite{Zel}.
The energy continuity equation follows from the Einstein equations
$$\dot\rho=-3(\rho+p)\frac{\dot{a}}{a}.$$
So, for the equation of state $\rho=p$, one is obtained the dependence $\rho\sim a^{-6}.$
The rigid state of matter can be formed by a free massless scalar field \cite{PZakh}.

Including executing fitting, we write the {\it conformal Friedmann equation} \cite{PP}
with use of significant conformal partial
parameters, discarding all other insignificant contributions
\begin{eqnarray}\label{Friedmannconf}
\left(\frac{dD}{d\eta}\right)^2&\equiv&
\left(\frac{{a}'}{a}\right)^2=\\
&=&\left(\frac{{\cal H}_0}{c}\right)^2\left[\Omega_{\rm rigid}\left(\frac{a_0}{a^4}\right)+
\Omega_{\rm M}\left(\frac{a_0}{a}\right)\right].\nonumber
\end{eqnarray}
In the right side of (\ref{Friedmannconf}) there are densities $\rho (a)$ with corresponding conformal weights; 
in the left side a comma denotes a derivative with respect to conformal time. The conformal Friedmann equation
ties intrinsic time interval with conformal time one. If we have accepted the intrinsic York's time \cite{York}
in Friedmann equations (\ref{Friedmannclass}), (\ref{Friedmannconf}), we should have lost the connection between 
temporal intervals\footnote{``The time is out of joint''. William Shakespeare. {\it Hamlet}. Act 1. Scene V. Longman, London (1970).}.
After introducing new dimensionless variable $x\equiv {a}/{a_0},$
the conformal Friedmann equation (\ref{Friedmannconf}) takes a form
\bea
&&\left(\frac{2c}{\sqrt{\Omega_{\rm M}} {\cal H}_0}\right)^2x^2\left(\frac{dx}{d\eta}\right)^2=\nonumber\\
&=&4x^3-g_3\equiv 4(x-e_1)(x-e_2)(x-e_3),\label{Weix}
\eea
where one root of the cubic polynomial in the right hand side (\ref{Weix})
is real, other are complex conjugated
\bea
e_1&\equiv& \sqrt[3]{\frac{\Omega_{\rm rigid}}{\Omega_{\rm M}}}\frac{1+\imath\sqrt{3}}{2},\qquad
e_2\equiv -\sqrt[3]{\frac{\Omega_{\rm rigid}}{\Omega_{\rm M}}},\nonumber\\
e_3&\equiv& \sqrt[3]{\frac{\Omega_{\rm rigid}}{\Omega_{\rm M}}}\frac{1-\imath\sqrt{3}}{2}.\nonumber
\eea
The invariants are the following
$$g_2=0,\qquad g_3= -\frac{4\Omega_{\rm rigid}}{\Omega_{\rm M}}.$$
where ${\cal H}_0$ is the conformal Hubble constant. The conformal Hubble parameter is defined via the Hubble parameter as
${\cal H}\equiv (a/a_0)H$.
The differential equation (\ref{Weix}) describes an effective problem of classical mechanics --
a falling of a particle with mass $8c^2/(\Omega_M{\cal H}_0^2)$ and
zero total energy in a central field with repulsive potential
$$U(x)=\frac{g_3}{x^2}-4x.$$
Starting from an initial point $x=0$ it reaches a point $x=1$ in a finite time $\eta_0$.
We get an integral from the differential equation (\ref{Weix})
\begin{equation}
\int\limits_{1/(1+z)}^1\frac{x dx}{\sqrt{4x^3-g_3}}=-\frac{\sqrt{\Omega_{\rm M}}{\cal H}_0}{2c}\eta.
\end{equation}

Then, we introduce a new variable $u$ by a rule
\begin{equation}\label{xpitau}
x\equiv\wp(u).
\end{equation}
Weierstrass function $\wp (u)$ \cite{Whittaker} satisfies to the differential equation
$$
\left[\frac{d\wp(u)}{du}\right]^2=4\left[\wp (u)-e_1\right]\left[\wp (u)-e_2\right]\left[\wp (u)-e_3\right],
$$
with
$$\wp (\omega_\alpha)=e_\alpha,\qquad \wp'(\omega_\alpha)=0,\qquad \alpha=1,2,3.$$
The discriminant is negative
$$\Delta\equiv g_2^3-27g_3^2<0.$$
The Weierstrass $\zeta$-function satisfies to conditions of quasi-periodicity
$$\zeta (\tau+2\omega)=\zeta (\tau)+2\eta,\qquad
\zeta (\tau+2\omega')=\zeta (\tau)+2\eta',$$
where
$$\eta\equiv\zeta (\omega),\qquad \eta'\equiv \zeta (\omega').$$

The {\it conformal age -- redshift relationship} is obtained in explicit form
\begin{equation}\label{age}
\eta=\frac{2c}{\sqrt{\Omega_{\rm M}}{\cal H}_0}\left(\zeta\left[\wp^{-1}\left(\frac{1}{1+z}\right)\right]-
\zeta\left[\wp^{-1}(1)\right]\right).
\end{equation}
Rewritten in the integral form the Friedmann equation is known in cosmology as the {\it Hubble law}.
The explicit formula for the {\it age of the Universe} can be obtained
\begin{equation}
\eta_0=\frac{2c}{\sqrt{\Omega_{\rm M}}{\cal H}_0}
\left(\zeta \left[\wp^{-1}(0)\right]-\zeta \left[\wp^{-1}(1)\right]\right).
\end{equation}
An interval of coordinate conformal distance is equal to an interval of conformal time $dr=d\eta$, so we can
rewrite (\ref{age}) as {\it conformal distance -- redshift relation}.

A relative changing of wavelength of an emitted photon corresponds to a relative changing of the scale
$$z=\frac{\lambda_0-\lambda}{\lambda}=\frac{a_0-a}{a},$$
where $\lambda$ is a wavelength of an emitted photon, $\lambda_0$ is a wavelength of absorbed photon.
The Weyl treatment \cite{PP} suggests also a possibility to consider
\begin{equation}\label{WeylCC}
1+z=\frac{m_0 a_0}{[a(\eta) m_0]},
\end{equation}
where $m_0$ is an atom original mass.
Masses of elementary particles, according to Conformal cosmology interpretation (\ref{WeylCC}), become running
$$m(\eta)=m_0 a(\eta).$$

The photons emitted by atoms of the distant stars billions of years ago, remember the size of atoms.
The same atoms were determined by their masses in that long time. Astronomers now compare the spectrum of
radiation with the spectrum of the same atoms on Earth, but with increased since that time.
The result is a redshift $z$ of Fraunhofer spectral lines.

In conformal coordinates photons behave exactly as in Minkowski space.
The time intervals $dt= - a dr$ used in Standard cosmology and the time interval used in Conformal
cosmology $d\eta = - dr$ are different.
The conformal luminosity distance $d_L(z)_{CC}$ is related to the standard luminosity one $d_L(z)_{SC}$ as
\cite{PZakh}
$$d_L(z)_{CC}= (1+z)d_L(z)_{SC}=(1+z)^2r (z),$$
where $r(z)$ is a coordinate distance.
For photons $dr/d\eta=-1,$ so we obtain the explicit dependence:
{\it luminosity distance -- redshift relationship}
\bea\label{r(z)conformal}
&&d_L(z)_{CC}=\\
&=&\frac{2c(1+z)^2}{\sqrt{\Omega_{\rm M}}{\cal H}_0}
\left(\zeta\left[\wp^{-1}\left(\frac{1}{1+z}\right)\right]-
\zeta\left[\wp^{-1}(1)\right]\right).\nonumber
\eea

{\it The effective magnitude -- redshift relation} in Conformal cosmology has a form
\begin{equation}\label{mMConf}
m(z)-M=5{\rm lg} [d_L(z)_{CC}]+{\cal M}.
\end{equation}

\section{Comparisons of approaches}

The Conformal cosmological model states that conformal quantities are observable magnitudes.
The Pearson $\chi^2$-criterium was applied in \cite{PZakh}
to select from a statistical point of view the best fitting of Type Ia supernovae data
\cite{Riess2004}.
The rigid matter component $\rho_{\rm rigid}$ in the Conformal model substitutes the $\Lambda$-term of the Standard model. 
It corresponds to a rigid state of matter, when the energy density is equal to its pressure.
The result of the treatment is: the best-fit of the Conformal model is almost the same quality approximation as
the best-fit of the Standard model.

\begin{figure}[tbp]
\begin{center}
\includegraphics[width=3.1in]{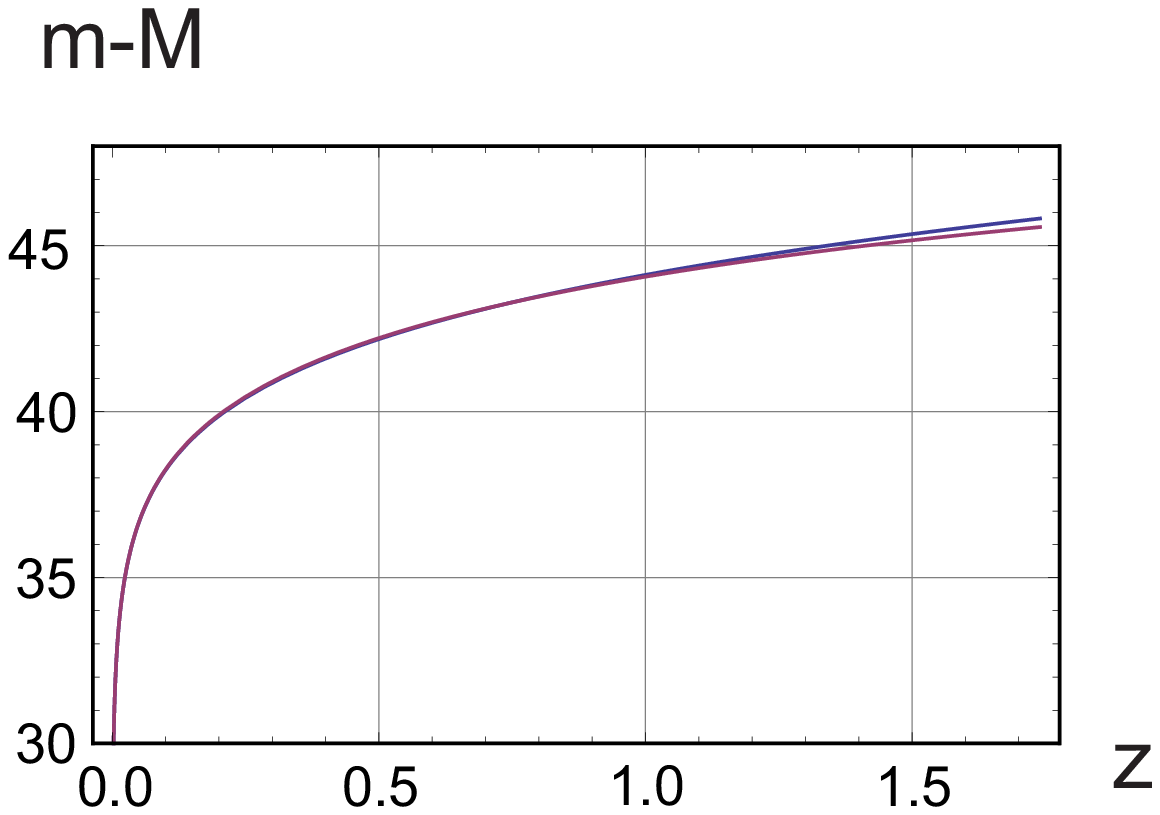}
\caption{\small Curves: the effective magnitude -- redshift relation of the two models.}
\label{Comparison}
\end{center}
\end{figure}
\begin{figure}[tbp]
\begin{center}
\includegraphics[width=3in]{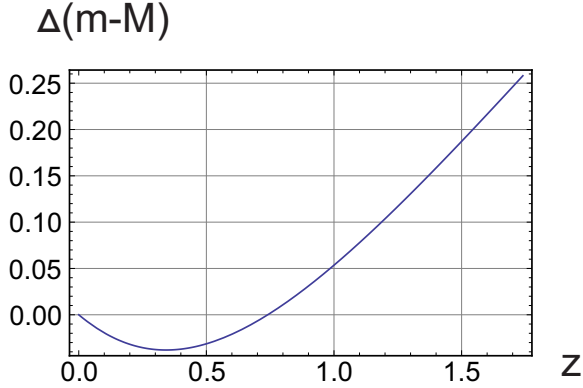}
\caption{\small Difference between curves of the two models: The effective magnitude -- redshift relation.}
\label{Delta}
\end{center}
\end{figure}

Curves of the two models are shown in Fig.\ref{Comparison}.
A fine difference between predictions of the models (\ref{mMConf}) and (\ref{mMcl}):
effective magnitude -- redshift relation
$$\Delta (m(z)-M)=5{\rm lg}[d_L(z)_{CC}]-5{\rm lg}[d_L(z)_{SC}]$$
is depicted in Fig.\ref{Delta}. 
The differences between the curves are observed in the early and in the past stages of the Universe's evolution.

In Standard cosmology the Hubble, deceleration, jerk parameters are defined as \cite{Riess2004}
\begin{eqnarray}
H(t)&\equiv&+\left(\frac{\dot{a}}{a}\right)=H_0\sqrt{\frac{\Omega_{\rm M}}{a^3}+\Omega_\Lambda},\nonumber\\
q(t)&\equiv&-\left(\frac{\ddot{a}}{a}\right)\left(\frac{\dot{a}}{a}\right)^{-2}=
\frac{\Omega_{\rm M}/2-\Omega_\Lambda a^3}{\Omega_{\rm M}+\Omega_\Lambda a^3},\nonumber\\
j(t)&\equiv&+\left(\frac{\dot{\ddot{a}}}{a}\right)\left(\frac{\dot{a}}{a}\right)^{-3}=1.
\end{eqnarray}
As we have seen, the $q$-parameter changes its sign during the Universe's evolution at an inflection point
$$a^{*}=\sqrt[3]{\frac{\Omega_{\rm M}}{2\Omega_\Lambda}},$$
the $j$-parameter is a constant.

We can define analogous parameters in Conformal cosmology also
\begin{eqnarray}
{\cal H}(\eta)&\equiv&+\left(\frac{{a}'}{a}\right),\label{Hparameter}\\
{q}(\eta)&\equiv&-\left(\frac{{a}''}{a}\right)\left(\frac{{a}'}{a}\right)^{-2},\label{qparameter}\\
{j}(\eta)&\equiv&+\left(\frac{{a}'''}{a}\right)\left(\frac{{a}'}{a}\right)^{-3}.\label{jparameter}
\end{eqnarray}
Let us calculate the conformal parameters with use of the conformal Friedmann equation (\ref{Friedmannconf}).
The Hubble parameter
$${\cal H}(\eta)=\frac{{\cal H}_0}{a^2}\sqrt{\Omega_{\rm rigid}+\Omega_{\rm M} a^3}>0;$$
the deceleration parameter
$$q(\eta)=\left(\frac{\Omega_{\rm rigid}-(\Omega_{\rm M}/2)a^3}{\Omega_{\rm rigid}+
\Omega_{\rm M}a^3}\right)>0,$$
so the scale factor grows with deceleration;
the jerk parameter
$$j(\eta)=\frac{3\Omega_{\rm rigid}}{\Omega_{\rm rigid}+\Omega_{\rm M}a^3}>0$$
changes from 3 to $3\Omega_{\rm rigid}$.
The dimensionless parameter $q(\eta)$ and $j(\eta)$ are positive during all evolution. 
The Universe has not been undergone a jerk.

\section{Conclusions}

Weierstrass and Jacobi functions traditionally used for a long time in classical mechanics and astronomy, 
are in demand in theoretical cosmology also. The conformal age -- redshift relation, 
and the effective magnitude -- redshift relations, that are basis formulae for observable cosmology, 
are expressed explicitly in meromorphic functions. Instead of integral relations, which are used to in cosmology, 
the derived formulae are expressed through higher transcendental functions, easy to use, 
because they are built-in analytical software package MATHEMATICA.

The Hubble Space Telescope cosmological supernovae Ia team presented data of high redshifts.
Classical cosmological and Conformal cosmological approaches fit the Hubble diagram with equal accuracy.
According to concepts of Conformal gravitation, conformal quantities of General Relativity are interpreted 
as physical observables. The conformal cosmological interpretation is preferable because of explaining the 
resent data without adding the $\Lambda$-term.

It is appropriate to remind the correct statement of the Nobel laureate in Physics Steven Weinberg
\cite{Three} about interpretation of experimental data on redshift. ``{\it I do not want to give the impression
that everyone agrees with this interpretation of the red shift.
We do not actually observe galaxies rushing away from us; all we are sure of is that the lines in their spectra
are shifted to the red, i. e. towards longer wavelengths. There are eminent astronomers who doubt that the
red shifts have anything to do with Doppler shifts or with expansion of the universe}''.

\section*{Acknowledgment}

For fruitful discussions I would like to thank Profs. A.B. Arbuzov, R.G. Nazmitdinov, and V.N. Pervushin.


\end{document}